**Subthreshold Resonances in the Scattering of μ^-Mesons by Atomic Nuclei**

F.M. Pen'kov[1,2,3], and A. Mukhametkaliuly[3,4,*]

[1]*Al-Farabi Kazakh National University, Almaty*
[2]*Institute of Applied Sciences and Information Technologies, Almaty*
[3]*Institute of Nuclear Physics, Almaty*
[4]*Joint Institute for Nuclear Research, Dubna*

Abstract

A theoretical study of pre-threshold resonant excitation of atomic nuclei is presented. A formalism for utilizing leptonic probes in the energy region of a Feshbach–Baz pre-threshold resonance is proposed. The resonance positions and widths are shown to be intrinsically linked to the nuclear structure parameters. The correspondence of the giant dipole resonance (GDR) region for heavy nuclei to low probe energies, as well as its large resonance widths, is demonstrated. Through this framework, free from the intense background, the observable signatures of GDR excitation and de-excitation channels, including nuclear fission, are predicted.

**Introduction**

In 1958, Feshbach [1] published a work on the theory of nuclear resonances, which laid the foundation for understanding a large number of resonant scattering processes arising from bound states in closed channels. Literally within a year, Baz' [2] published a work on subthreshold resonances in the scattering of charged particles, in which subthreshold resonances form an infinite series accumulating toward the excitation threshold of the subsystem, reflecting a hydrogen-like spectrum both in energy and in multipolarity. In Ref. [2], an analytical expression for the S-matrix of elastic scattering was obtained, and the broad applicability of the results of that work can be seen in the review [3]. We note that the effects described in Ref. [2] were also used in elementary-particle physics to explain the anomaly in the scattering of $\pi^{+}$mesons by atomic nuclei [4].

Since Baz' resonances (BRs) reflect a hydrogen-like spectrum, the greater the mass of the incident particle, the farther the resonances lie from the threshold and the greater the spacing between neighboring resonances. One manifestation of such effects is the scattering of negative muons by atomic nuclei. The absence of strong interaction and the sufficiently well-studied spectra of mesoatoms [5], in principle, make it possible to use beams of negative muons with energies up to about ten MeV to study electromagnetic transitions in nuclei. In this case, when the nucleus is excited to energies at which nuclear reactions are allowed, one can consider the possibility of studying such nuclear transitions, up to and including fission, by analyzing the widths and positions of subthreshold BRs.

Modern meson factories make it possible to achieve substantial muon currents at relatively low muon kinetic energies. Thus, on the μE4 channel of the high-current proton accelerator at the Paul Scherrer Institute (PSI), positive-muon currents reach $6.8 \cdot 10^{8}$ muons/s in the energy range 0.5÷7.5 MeV [6]. At the same time, the muon energy resolution reaches $10^{-4}$ [2]. Moreover, tasks requiring a resolution of $\sim 3 \cdot 10^{-5}$ have been identified (see, for example, [7]). Here we consider the scattering of negative muons, for which the beam parameters are significantly inferior to those of positive-muon beams, but it is precisely the record-scale beam parameters $\mu^+$ that we have chosen to estimate the possibilities of observing resonance phenomena in both elastic and inelastic processes on negative-muon channels. Previously, the authors published a note [8] on the possibility of observing resonances in elastic scattering of negative muons by nuclei, based on calculations of the widths in the plane-wave approximation for zero-radius nuclei, which is poorly suited for analyzing scattering in the range accessible to accelerator technology. In addition, the E1-transition widths were estimated from dimensional considerations within the Weisskopf model, which proved to be too crude an estimate for planning experimental studies of BRs with muon beams. Therefore, to increase the reliability of the conclusions, in the present work we used only experimental values of E1-transition probabilities in nuclei, including giant dipole resonances. Accordingly, we went beyond the plane-wave approximation and the model of pointlike atomic nuclei.

* Contact author: adilm2306@gmail.com

## Theoretical Framework

Below, we consider the scattering of $\mu^-$ mesons of mass $m$ by atomic nuclei with mass number $A$ and charge number $Z$. The potential of the muon–nucleus interaction is determined solely by the Coulomb interaction with pointlike protons:

$$U = -e^2 \sum_{i=1}^{Z} \frac{1}{|\boldsymbol{\rho} - \boldsymbol{r}_i|}. \qquad (1)$$

Here the coordinates $\boldsymbol{r_i}$ and $\boldsymbol{\rho}$ are the coordinates of the protons and the muon relative to the center of mass of the nucleus, and $e$ is the elementary charge. According to the BR model, the system consisting of a nucleus in the ground state with energy $\varepsilon_0$ and a negative muon with relative-motion energy $E_k = \frac{\hbar^2 k^2}{2m}$ can have an energy equal to that of the "mesoatom with an excited nucleus" system, with energy $E_k + \varepsilon_0 = E_c + \varepsilon_1$. Here $\varepsilon_1$ is the binding energy of the nucleus in the excited state, and $E_c$ is the binding energy of the mesoatom. Let us denote the wave functions of these states by $\Psi_{\boldsymbol{k}}$ and $\Psi_b$, respectively. If potential (1) provides coupling between the states $\Psi_{\boldsymbol{k}}$ and $\Psi_b$, then, in scattering at the muon kinetic energy

$$E_{res} = E_c + \varepsilon_1 - \varepsilon_0$$

a resonance is observed, whose width can be written within the Feshbach approach [1]:

$$\Gamma = \frac{mk}{\pi\hbar^2} \int |\langle \Psi_b | U | \Psi_{\boldsymbol{k}} \rangle|^2 \frac{d\Omega_k}{4\pi}. \qquad (2)$$

The wave functions in Eq. (2) satisfy the normalization conditions: $\langle \Psi_b | \Psi_b \rangle = 1$ $\langle \Psi_{\boldsymbol{k}} | \Psi_{\boldsymbol{k}'} \rangle = (2\pi)^3 \delta(\boldsymbol{k} - \boldsymbol{k}')$. Within the same approach, the coupling between the channels is assumed to be weak, and the wave functions $\Psi_{\boldsymbol{k}}$ and $\Psi_b$ are factorized: $\Psi_{\boldsymbol{k}} = \psi_{\boldsymbol{k}}^+(\boldsymbol{\rho}) \varphi_0(\boldsymbol{r_1}, \boldsymbol{r_2}, \dots, \boldsymbol{r_Z})$, $\Psi_b = \psi_\mu(\boldsymbol{\rho}) \varphi_1(\boldsymbol{r_1}, \boldsymbol{r_2}, \dots, \boldsymbol{r_Z})$ into the wave functions of the nucleus in the ground $\varphi_0$ and excited $\varphi_1$ states and the relative-motion wave functions $\psi_\mu(\boldsymbol{\rho})$ of the mesoatom and of the muon scattering state $\psi_{\boldsymbol{k}}^+(\boldsymbol{\rho})$ with scattering amplitude $f$, which will determine the cross section for muon scattering by the nucleus.

In the present work, only dipole coupling between the channels is considered

$$\langle \varphi_1 | U | \varphi_0 \rangle = \left\langle \varphi_1 \middle| e \frac{\boldsymbol{d\rho}}{\rho^3} \middle| \varphi_0 \right\rangle, \qquad (3)$$

which implies an electromagnetic $E1$ transition of the nucleus to an excited state. In this case, the partial angular components of the functions $\psi_\mu(\boldsymbol{\rho})$ and $\psi_{\boldsymbol{k}}^+(\boldsymbol{\rho})$ must also correspond to orbital angular momenta differing by 1. Since we consider the resonance farthest from the nuclear excitation threshold, which occurs at the largest value of the mesoatom binding energy, $\psi_\mu(\boldsymbol{\rho})$ must correspond to the ground state of the mesoatom in the S wave. Accordingly, integral (2) will be determined by only one value of the partial harmonic $\psi_{\boldsymbol{k}}^+(\boldsymbol{\rho})_l$ with $l = 1$. In other words, the scattering of a negative muon in the P wave by a nucleus is considered immediately below the threshold for its excitation to a state that provides a$E1$ transition to the nuclear ground state.

Because of the coupling between the channels (3), the wave functions $\psi_{\boldsymbol{k}}^{+}(\boldsymbol{\rho})$ and $\psi_{\mu}(\boldsymbol{\rho})$ will have a correction of the scale of (3). The use of such functions would clearly exceed the accuracy implied by using Eq. (2) with coupling (3). Therefore, within the approach, the radial functions of the angular projections $\psi_{k}^{+}(\boldsymbol{\rho})$ and $\psi_{\mu}(\boldsymbol{\rho})$ satisfy the Schrödinger equations without taking coupling (3) into account:

$$\begin{aligned} -\frac{\hbar^2}{2m}\Delta_l R_{kl} + \langle\varphi_0|U|\varphi_0\rangle R_{kl} = E_k R_{kl}, \\ -\frac{\hbar^2}{2m}\Delta_l R_{\mu l} + \langle\varphi_1|U|\varphi_1\rangle R_{\mu l} = E_c R_{\mu l}. \end{aligned} \qquad (4)$$

Here $\Delta_l$ denotes the radial part of the three-dimensional Hamiltonian with orbital angular momentum $l$. Potentials of the type $\langle\varphi_0|U|\varphi_0\rangle$ and $\langle\varphi_1|U|\varphi_1\rangle$ determine the Coulomb potential of the nucleus. At distances smaller than the characteristic nuclear size $R$, they represent the internal potential of the charged region, which is determined by the proton distribution in the nucleus and is subsequently assumed to be uniform in the calculations. At distances larger than the nuclear size, these potentials reduce to the usual Coulomb potential: $\langle\varphi_i|U|\varphi_i\rangle \xrightarrow{\rho\gg R} -\frac{Ze^2}{\rho}$.

After expanding $\psi_{\boldsymbol{k}}^{+}(\boldsymbol{\rho})$ and $\psi_{\mu}(\boldsymbol{\rho})$ in angular harmonics, Eq. (2) can be written as

$$\Gamma_{el} = \frac{4}{3}\frac{me^2\boldsymbol{d}_{10}^2 k}{\hbar^2}\left|\int R_{\mu 0}(\rho)\, R_{k1}(\rho) d\rho\right|^2 \delta_{l,1}\,, \qquad (5)$$

where, for definiteness, the width $\Gamma_{el}$ is defined as the resonance width in the presence of elastic processes only, $\boldsymbol{d}_{10}^2 = (\langle\varphi_1|\boldsymbol{d}|\varphi_0\rangle, \langle\varphi_0|\boldsymbol{d}|\varphi_1\rangle)$ is the square of the dipole transition, the functions $R_{\mu 1}(\rho)$ are normalized only over the radial part of the volume, and for the continuous spectrum a dimensionless asymptotic form is chosen:

$$\int_0^\infty R_{\mu 0}^2 \rho^2 d\rho = 1; \qquad R_{kl} \xrightarrow{\rho\to\infty} \frac{\sin\left(k\rho - \frac{l\pi}{2} + \delta_l\right)}{k\rho}.$$

Here $\delta_l$ determines the phase of elastic scattering by the potential $\langle\varphi_0|U|\varphi_0\rangle$ with orbital angular momentum $l$, while the Kronecker symbol $\delta_{l,1}$ in (5) emphasizes that resonant scattering for the transition operator (3) exists only in the P wave of elastic scattering. Thus, for elastic scattering through an angle $\vartheta$ without the possibility of an inelastic transition, the resonant scattering amplitude, within the Feshbach approach and according to Eq. (5), can be written in the form of a Breit–Wigner amplitude:

$$f_{res} = -3\, P_1(\cos(\vartheta))\frac{1}{2k}\frac{\Gamma_{el}}{E - E_{res} + \frac{i\Gamma_{el}}{2}}, \qquad (6)$$

which gives the resonant elastic-scattering cross section:

$$\sigma_{res} = \frac{3\pi}{k^2}\frac{\Gamma_{el}^2}{(E - E_{res})^2 + \frac{\Gamma_{el}^2}{4}}.$$

The mechanisms of resonance formation through the formation of a mesoatom with an excited nucleus allow a reaction channel in which the excited nucleus emits a gamma quantum with energy$E_\gamma = \varepsilon_1 - \varepsilon_0$, transferring the system to the state of a mesoatom with the nucleus in its ground state. Therefore, the elastic processes are described by the total width $\Gamma = \Gamma_{el} + \Gamma_\gamma$, with radiative width $\Gamma_\gamma$:

$$\sigma_{res}^{ell} = \frac{3\pi}{k^2}\frac{\Gamma_{el}^2}{(E - E_{res})^2 + \frac{\Gamma^2}{4}}. \qquad (7)$$

Accordingly, the inelastic processes are determined by the cross section for gamma-quantum emission in resonant scattering of negative muons by atomic nuclei:

$$\sigma_{res}^{\gamma} = \frac{3\pi}{k^2}\frac{\Gamma_{el}\Gamma_\gamma}{(E - E_{res})^2 + \frac{\Gamma^2}{4}}. \qquad (8)$$

The values $\Gamma_\gamma$ for $E1$ transitions can be taken from any database (see, for example, [9]). In this case, it is reasonable to eliminate $\boldsymbol{d}_{10}^2$ from Eq. (5), replacing it by $\Gamma_\gamma$ using the well-known formula (see, for example, [10]):

$$\Gamma_\gamma = \frac{4E_\gamma^3}{3\hbar^3 c^3}\,|\boldsymbol{d}_{10}|^2 \qquad (9)$$

Equations (5) and (9) imply the relation between the width of elastic resonant muon scattering by a nucleus and the radiative width $\Gamma_{\mathrm{el}} = \eta\Gamma_\gamma$:

$$\eta = \frac{c^3 me^2\hbar k}{E_\gamma^3}\left|\int R_{\mu 0}(\rho)\, R_{k1}(\rho) d\rho\right|^2. \qquad (10)$$

The quantity $\eta$ is calculated numerically for any specified nuclear charge density, whereas for pointlike nuclei the calculation can be performed analytically. Then, using the quantity $\eta$, the maximum values of the cross sections (7) and (8) at $E = E_{res}$ can be written as:

$$\sigma^{ell}_{res,\max} = \frac{12\pi}{k^2}\frac{\eta^2}{(1+\eta)^2}, \qquad \sigma^{\gamma}_{res,\max} = \frac{12\pi}{k^2}\frac{\eta}{(1+\eta)^2}. \quad (11)$$

Considering the cross sections (11) as a function of the quantity $\eta$, one can see that the maximum of the inelastic cross sections is reached at $\eta = 1$, i.e., at $\Gamma_{\mathrm{el}} = \Gamma_\gamma$, and amounts to one quarter of the unitarity limit $\sigma_U = \frac{12\pi}{k^2}$, coinciding with the elastic-process cross section. The calculation presented below shows that $\eta$ can be both much smaller than 1 and much larger than 1.

For further analysis, let us define the muonic Bohr radius for a nucleus with charge number Z: $a_z = \frac{\hbar^2}{Ze^2 m}$.. Its numerical value, $a_z = \frac{255.94}{\mathrm{Z}}$ fm, indicates that the condition $R \approx a_z$ is satisfied for the middle of the periodic table at $Z = 45 \div 46$. In the case of light elements, the nuclei can be regarded as pointlike by replacing the potentials in (4) with their asymptotic values $-\frac{Ze^2}{\rho}$. Therefore, the matrix element can be calculated analytically:

$$\eta = \frac{8m\pi c^3 e^2 \hbar}{E_\gamma^3 a_Z^2}\frac{\exp\left(-4\frac{\arctan(a_Z k)}{a_Z k}\right)}{(a_Z^2 k^2 + 1)\left(1 - \exp\left(-2\frac{\pi}{a_Z k}\right)\right)}. \quad (12)$$

The asymptotic behavior of expression (12) at $a_Z k \gg 1$:

$$\eta \xrightarrow{k\to\infty} \frac{4m\pi c^3 e^2 \hbar}{E_\gamma^3 a_Z^3 k}$$

shows a decrease in the coefficient $\eta$ at high resonance energies. By contrast, at $a_Z k \ll 1$, expression (12) tends to a constant which, taking into account $|E_c| \approx E_\gamma$ in the same limit, is expressed in terms of the fundamental constants and the nuclear charge number:

$$\eta \xrightarrow{k\to 0} \frac{64\pi \exp(-4)}{\alpha^3 Z^4}. \quad (13)$$

where $\alpha^3$ is the fine-structure constant. For Z<55, i.e., in the range of applicability of (12), the quantity $\eta$ will be greater than 1. Thus, $\Gamma_{\mathrm{el}}$ can be both significantly greater than $\Gamma_\gamma$ at low resonance energies $\left(|E_c| \approx E_\gamma\right)$ and smaller at $|E_c| \ll E_\gamma$.

Further calculations were performed taking the nuclear charge density into account. The nucleus was assumed to be a uniformly charged sphere with a Coulomb potential of radius $R = 1.2\, A^{\frac{1}{3}}\, fm$, the same for both the ground state of the nucleus and its excited state, i.e.,$\langle\varphi_0|U|\varphi_0\rangle = \langle\varphi_1|U|\varphi_1\rangle$:

$$\langle\varphi_0|U|\varphi_0\rangle = -Ze^2 \begin{cases} \frac{3}{2R} - \frac{\rho^2}{2R^3}: & \rho < R \\ \frac{1}{\rho}: & \rho \geq R. \end{cases} \quad (14)$$

For potential (14), all solutions of Eq. (4), both in the internal region ($\rho < R$)and in the external region ($\rho \geq R$), are expressed analytically in terms of confluent hypergeometric functions (see, for example, [11]); matching the solutions at $\rho = R$ determines both the mesoatom spectrum and the wave functions of the discrete and continuous spectra. The only equation that cannot be solved analytically is the transcendental equation for the mesoatom spectrum, which was always solved numerically. In addition, integration of the wave functions from (10) over the internal and external regions also has no analytical representation and was performed numerically.

**Results and Discussion**

The calculation results within the model described above are presented in Table

1. The same table also presents experimental data on the basis of which the prospects for studying resonant scattering of negative muons by atomic nuclei can be assessed. Table 1 gives the coefficient values $\eta = \Gamma_{ell}/\Gamma_{\gamma}$, which determine the maximum values of the cross sections of the elastic and inelastic processes (11), and the values of the total cross section $\sigma_{tot,max}$ at the resonance energy. It can be seen that the calculated values $\eta$ can be both less than 1 and substantially greater. In addition, the table gives the ratio $(\Gamma_{ell} + \Gamma_{\gamma})/E_{res}$, which makes it possible to assess the fundamental possibility of observing resonant scattering at existing experimental facilities. Transitions with high resonance energies were excluded after analysis because their widths correspond not to EM transitions but to open nuclear channels with the emission of several neutrons. These transitions are of interest for studying nuclear structure, but no data are available for their$\Gamma_{\gamma}$ values, which, in turn, does not allow the elastic-scattering width and, accordingly, the total cross sections to be calculated. The table selectively lists nuclear isotopes for which the widths of $E1$ transitions to the nuclear ground state are known. The purpose of the selection is to show different values of $\eta$ and $E_{res}$ for different mass ranges of stable nuclei. The short-lived $^{11}$Be is retained in the table to demonstrate the record value of $\eta$. The last column gives the values of the total cross section $\sigma_{EM,max} = \frac{12\pi}{k^2}\,\eta/(1+\eta)$ of electromagnetic transitions at the resonance energies. It can be seen that these cross sections can reach tens and even hundreds of barns. Nevertheless, the calculations show that the relative width of all single-particle $E1$ transitions is so small that it does not allow

Table 1. Experimental data and calculated values of resonance parameters in the scattering of $\mu^{-}$mesons by atomic nuclei. Here $E_{\gamma}$ and $\Gamma_{\gamma}$ are the experimental values of the energy and width of $E1$ transitions to the nuclear ground states [9], and $E_c$ is the calculated energy of the mesoatom with an excited nucleus, $\Gamma_{EM}$ $=(\Gamma_{el} + \Gamma_{\gamma})$. The remaining quantities are defined in the text.

| Nucleus | $E_{\gamma}$, MeV | $\Gamma_{\gamma}$, eV | $E_c$, MeV | $E_{res}$, MeV | $\eta_{calc}$ | $\Gamma_{el}$, eV | $\frac{\Gamma_{EM}}{E_{res}}$ | $\sigma_{EM,max}$ |
|---|---|---|---|---|---|---|---|---|
| $^{11}$Be | 0.32 | 0.0039 | 0.0450 | 0.275 | 125.5 | 0.500 | $1.82\cdot 10^{-6}$ | 251 |
| $^{48}$Ti | 3.699 | 0.0148 | 1.272 | 2.4271 | 1.572 | 0.023 | $1.57\cdot 10^{-8}$ | 17.5 |
| $^{92}$Zr | 3.638 | 0.0476 | 3.636 | 0.0022 | 3.127 | 0.149 | $8.93\cdot 10^{-5}$ | 23900 |
| $^{138}$Ba | 6.193 | 1.8249 | 6.120 | 0.073 | 0.815 | 1.488 | $4.54\cdot 10^{-5}$ | 427 |
| $^{146}$Nd | 7.163 | 0.9485 | 6.791 | 0.372 | 0.561 | 0.533 | $3.98\cdot 10^{-6}$ | 67.1 |

the scattering resonances to be observed at existing facilities with muon channels.

At present, resonant scattering with excitation of the giant dipole resonance (GDR) in heavy nuclei may be of experimental interest. In this case, the values of the muon resonance energies are relatively low and do not allow the nuclei to be excited by direct Coulomb excitation. A separate issue is the values of the radiative widths $\Gamma_{\gamma 0}$ from collective excitations with a broad spectrum of decay channels, including nuclear ones. The sum of the widths of all channels (neutron emission, fission) amounts to several MeV, and against this background direct measurement of the radiative width is extremely difficult and is not reported in databases. Moreover, the physics of the collective-excitation process implies a sum of individual dipole transitions whose amplitudes and cross sections are not directly described by a Breit–Wigner function for an isolated state. The modern scheme for describing resonant photoabsorption operates with the concepts of average radiative widths $\Gamma_{\gamma 0}$ within the Standard Lorentzian model (SLO) or its extensions (see, for example, the description of database [12] in Ref. [13]). In such approaches, one can use the relation between the radiative width and the integral of the total photoabsorption cross section, which, in turn, can be calculated in any Lorentzian model, for example, in the form [14]. Using this estimation scheme gives:

$$\Gamma_{\gamma 0} = \frac{E_{GDR}^2 \Gamma}{12\pi\, c^2 \hbar^2} \sigma_{max} \,. \qquad (15)$$

Here $E_{GDR}$, $\Gamma$ and $\sigma_{max}$ are the GDR resonance energy, its width, and the maximum cross section, which are presented in database [12]. One of the experimental data sets is the set considered in [14]. It is this set that is used in Table 2 to estimate the average values $\Gamma_{\gamma 0}$. Since the widths of the resonant interaction of negative muons with nuclei (5) are determined by the same dipole transitions as the photoabsorption processes in the GDR, the relation between the width of the elastic muon-scattering channel and the average radiative width will be the same as for an isolated transition in Eq. (10).

Table 2. Experimental data and calculated values of resonance parameters for the scattering of $\mu^-$mesons by atomic nuclei in the GDR region. $\Gamma$ is the total photoabsorption width [14], and $\Gamma_{\gamma 0}$ is the radiative-transition width estimated from Eq. (15). The remaining quantities are defined above.

| Nucleus | $E_{GDR}$, MeV | $\Gamma$, MeV | $E_c$, MeV | $E_{res}$, MeV | $\Gamma_{\gamma 0}$, keV | $\eta_{calc}$ | $\Gamma_{el}$, keV | $\frac{\Gamma_{EM}}{E_{res}}$ | $\sigma_{EM,max}$, barn |
|---|---|---|---|---|---|---|---|---|---|
| $^{232}$Th | 13.87 | 4.77 | 11.724 | 2.146 | 28.1 | 0.122 | 3.4 | 0.015 | 3.52 |
| $^{235}$U | 13.96 | 4.75 | 12.077 | 1.883 | 29.0 | 0.123 | 3.6 | 0.017 | 4.04 |
| $^{238}$U | 13.80 | 5.13 | 12.049 | 1.751 | 30.6 | 0.127 | 3.9 | 0.020 | 4.47 |

It should be noted that heavy nuclei are generally deformed and have two observable GDR excitation energies. The table gives the GDR excitation energy that corresponds to positive muon energies, i.e., the higher excitation energy.

The relative width $\frac{\Gamma_{EM}}{E_{res}}$ given in the table shows that the electromagnetic GDR decay channels can be observed even with a beam energy resolution of several percent. The last column of the table gives the maximum values of the total cross sections of the electromagnetic channel. It can be seen that, owing to the low values of the resonance energy, these cross sections reach several barns. Naturally, if the decay width of the muonic resonance coincides with the GDR decay width, then the observed EM cross sections will be $\Gamma_{\gamma 0}/\Gamma$, i.e., approximately $10^{-2}$ times smaller. However, even such cross sections make it possible to use existing beams of $\mu^-$mesons to observe photons accompanying GDR de-excitation through the EM channel. Unlike methods of GDR excitation by bremsstrahlung or even monochromatic quanta in modern-type facilities [15], muon beams are not accompanied by background gamma quanta, and all detected gamma quanta can be attributed only to quanta accompanying GDR decay. Thus, in such experimental arrangements, the true width of gamma transitions in the GDR can be determined directly, without reconstruction from data on the products of nuclear decay channels.

**Conclusions**

The theoretical consideration of a subthreshold resonance in the scattering of $\mu^-$mesons by heavy nuclei makes it possible to conclude that direct observation of gamma quanta accompanying GDR decay is possible. An experiment performed in such a configuration will provide more detailed information on the actual widths of EM transitions in collective nuclear excitation modes, narrowing the model descriptions of the formation and decay of collective states in atomic nuclei.

**Acknowledgments.** This work was supported by grant AP23486501 of the Ministry of Science and Higher Education of the Republic of Kazakhstan.

**Declaration of Generative AI in Scientific Writing:** During the preparation of this manuscript, the authors used ChatGPT, version 4o to translate the initial draft from Russian into English and polish the academic phrasing. The authors provided specific prompts to ensure standard physics terminology aligned with APS guidelines. After the automated translation, the authors thoroughly reviewed, edited, and verified the entire text to guarantee scientific accuracy, proper formatting of equations, and the integrity of the work. The authors assume full accountability for the final content of the manuscript.